\documentclass[english,aps,prstper,reprint,showpacs,titlepage,longbibliography]{revtex4-2}   

\usepackage[T1]{fontenc}
\usepackage[latin9]{inputenc}
\usepackage{geometry}		
\usepackage{graphicx}
\usepackage[above,below]{placeins}	
\usepackage{times}
\usepackage{hyperref}  
\hypersetup{colorlinks=true,urlcolor=blue,citecolor=blue,linkcolor=blue}   
\usepackage{enumitem}        
\setlist{nosep}              

\usepackage{todonotes}

\begin{document}

\begin{titlepage}

  \title{Development of the RIOT 2 for use in LA-supported classrooms}

  \author{Gina M. Quan, Cassandra Paul, Nithin Keshavamurthy, Owen Knight, Aarav Ghai, Kamalbabu Rupanagudi}
  \affiliation{Physics \& Astronomy, San Jos\'e State University, One Washington Square, San Jose, CA, 95192} 


  \begin{abstract}
This paper describes the development of the RIOT 2, an observational tool that can be used to study classrooms in which there are multiple members of an instructional team. The RIOT 2 is based on the Real-time Instructor Observational Tool (RIOT), which characterizes the behavior of an instructor over time. While the RIOT was previously developed for classrooms with only one instructor, our team has developed a version to observe classrooms supported by Learning Assistants (LAs). LA are undergraduate peer educators who facilitate discussions in active-learning classrooms. Unlike the RIOT, the RIOT 2 allows for the categorization of LA-instructor interactions. In this paper, we describe the development of RIOT 2 and present some initial data collected using the RIOT 2. While our data are preliminary, they illustrate how the RIOT 2 may be useful for future research and practice. This work is part of a broader research effort to understand how LA-faculty partnerships impact what LAs and instructors do in the classroom.
    \clearpage
  \end{abstract}

  \maketitle
\end{titlepage}

\section{Introduction}
The Learning Assistant (LA) Model incorporates undergraduate peer educators into active-learning university classrooms to facilitate group work and promote student engagement \cite{otero2006responsible}. While there exist variations in implementations of the model across different institutions, LA programs typically include three components: pedagogical training, where LAs learn about pedagogy; content preparation, where LAs and faculty partners discuss how to support students in their context; and practice, where LAs facilitate in-class discussions \cite{otero2006responsible}. LAs have been shown to support improvements in student learning, student motivation, and graduation rates \cite{tedeschi2023improving,kramer2023establishing,barrasso2021scoping,alzen2018logistic}. 

Even though the LA Model has been widely successful, additional research suggests significant variation in the model's implementation. Pedagogical training can take many forms across institutions, emphasizing different content and practices \cite{barrasso2021scoping,quan2017designing}. Within a single institution, LA-faculty partnerships can vary in the distribution of power in decision making---in some cases reflecting shared governance and others appearing more instructor-controlled \cite{davenport2018exploring}. In addition, LAs can take on different kinds of roles within their partnerships (e.g., co-instructors, informants, consultants) at different times \cite{indukuri2022characterizing,jardine2020positioning}. 

Our project is interested in how these differences in LA-faculty partnerships may also be connected to variations in what LAs and instructors do in the classroom. Toward this end, we have been developing and refining an observational protocol to characterize what LAs and instructors do in LA-supported classrooms. This development paper has two major aims: 
\begin{enumerate}
    \item Describe the development of the RIOT 2, a protocol to observe LA-supported classrooms.
    \item Present some initial data collected with the RIOT 2, to illustrate the potential of this tool.
\end{enumerate}
\section{Background: RIOT}

Observational protocols have been extensively used in education research to document and quantify what is happening in the classroom---a notable example example being the classroom observation protocol for undergraduate STEM (COPUS) \cite{smith2013classroom}. There exist a number of observational protocols, each developed to identify specific features, including the presence of different behaviors, what is happening in time, what instructors do, and what students do \cite{madsen2019resource, smith2013classroom, hora2015toward}. 

Our research focuses on the Real-time Instructor Observational Tool (RIOT), initially developed at UC Davis \cite{west2013variation}. The RIOT allows observers to continuously record what instructors are doing over time, resulting in a quantifiable breakdown of time spent on different instructor behaviors. This differs from other observational protocols which identify what students are doing, or those which capture the presence (but not duration) of specific behaviors.
\begin{figure}
  \includegraphics[width=\linewidth]{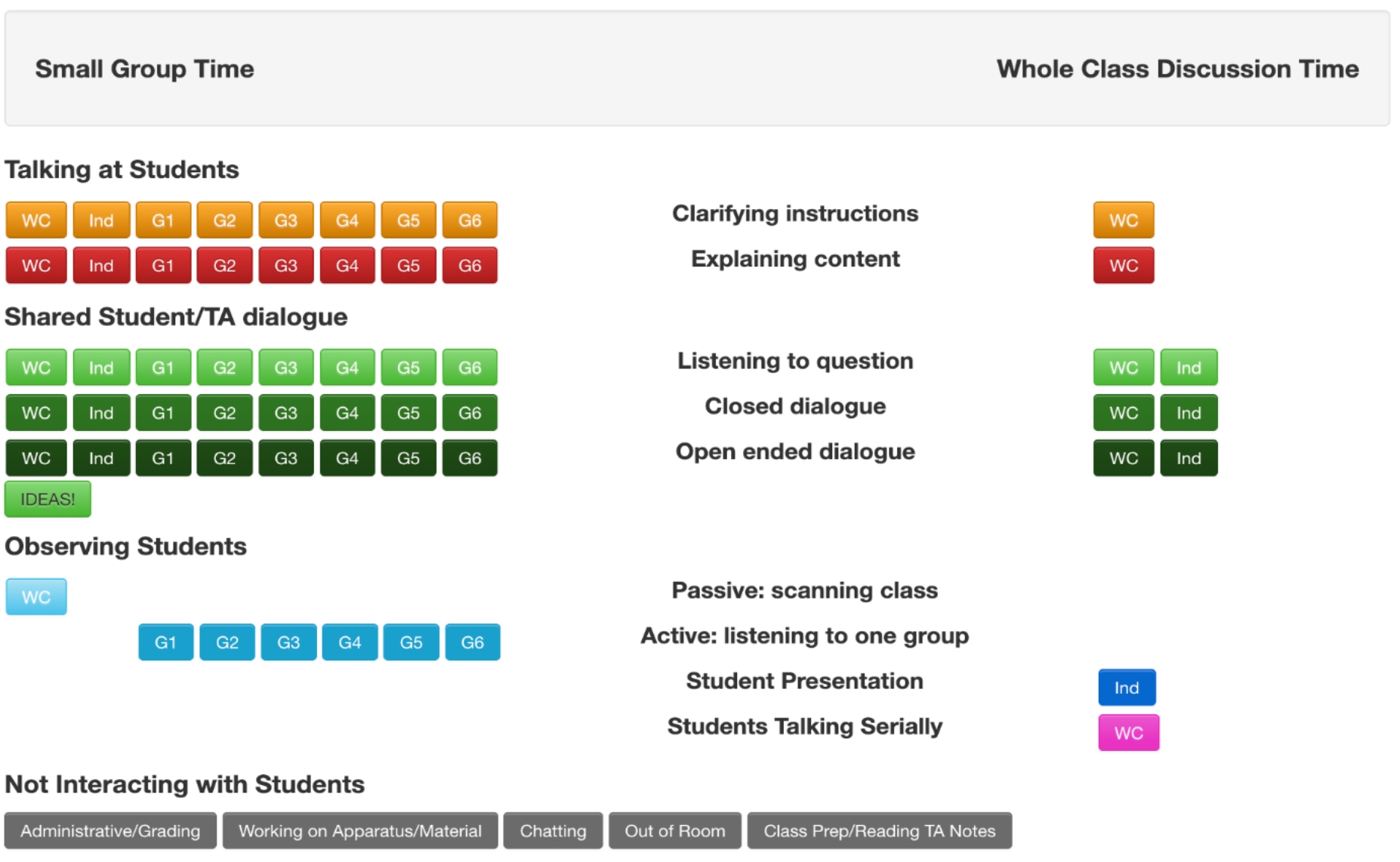}
  \caption{The original RIOT interface. Rows/colors correspond to instructor behaviors. \label{fig:riotinterface}}
\end{figure}

To use the RIOT, a trained observer attends a classroom session with the RIOT application on a tablet or computer. Figure \ref{fig:riotinterface} depicts the RIOT interface, and each of the colors/rows corresponds to a type of behavior. The columns describe which student(s) the instructor is interacting with---the whole class (WC), an individual student (IND), Group 1 (G1), etc. Buttons on the left hand side are to be used for times when the class is in small-group discussions, whereas the right hand side is for whole-class discussions. As the instructor engages in behaviors, the observer can click a button corresponding to that behavior. The RIOT interface assumes the instructor is engaging in that behavior until another button is clicked. Following the conclusion of a RIOT session, the application generates a plot showing the timeline of various instructor behaviors as well as pie charts depicting the allocation of time spent in different behaviors. 
\begin{table*}
    \centering
    \begin{ruledtabular}
        \begin{tabular}{p{2.5cm} p {4cm} p{11cm}}
        \textbf{Type} & \textbf{Category of Interaction} & \textbf{Description (Instructor is...)} \\
        \hline
        Talking at students & 
        Clarifying instructions & 
        Giving or clarifying classroom instructions, covering logistical issues, transitioning\\
        & Explaining content &
        Explaining physics concepts, answers, or processes. \\
        \hline
        Dialoguing with & 
Listening to question  & 
Listening to a student's question. \\
students & Engaging in closed dialogue & 
Asking a series of short questions meant to lead the student to a correct answer. Student contributions are typically short. \\
& Engaging in open dialogue & 
Discussing ideas dialogically. Students are contributing complete sentences. \\
& Ideas being shared & 
Participating in student-led conversation. Student contribution is complete sentences with concepts being challenged and worked on. \\ \hline

Observing students & 
Passive observing & 
Scanning room and assessing student progress from afar or browsing whiteboard work of groups for less than 10 seconds at a time. \\
& Active observing & 
Actively listening, focused on small groups or individuals. \\
& Students presenting & 
Listening to students presenting their work to the class. \\
& Students talking serially & 
Listening to students talking one-after-another. \\ \hline
Not interacting & 
Administrative and/or grading & 
Grading student homework, or discussing quizzes or other course policies. \\
& Class preparation/reading notes & 
Reading notes, or writing something on the board. \\
& Chatting & 
Chatting socially with students. This is not an interaction discussing physics. \\
& Working on apparatus and/or material & 
Helping students with experimental apparatus or computers without discussing physics content. \\
& Out of room & 
Left the room. 

        \end{tabular}
    \end{ruledtabular}
    \caption{Summary of instructor behaviors categorized by RIOT, adapted from Paul \& West \cite{paul2018using}}
    \label{tab:riotbehaviors}
\end{table*}

The RIOT includes 15 instructor behaviors which can be categorized into 4 different types. These are presented in Table 1 (adapted from \cite{paul2018using}). As a tool, the RIOT has been shown to be helpful for a number of uses, including identifying variations in implementations of a specific curriculum \cite{west2013variation}, supporting instructor reflection \cite{paul2018using}, measuring the buy-in of graduate TAs in reformed courses \cite{wilcox2016quicker}, and measuring the impact of TA professional development \cite{doucette2020professional}.

Given these use cases, our project decided to adapt the RIOT to be compatible with classrooms with Learning Assistants. As the RIOT was developed for use in single-instructor classrooms, it has primarily been used in those settings (with one notable exception \cite{odden2023implementing}). However, the RIOT lacks the ability to capture interactions between members of the instructional team (e.g., LAs observing instructor, LAs talking to instructor). Because these LA-faculty interactions can be substantial, meaningful behaviors, our project engaged in a process of iteratively developing and testing the RIOT 2, a RIOT-based protocol that accounts for classrooms where there are multiple members of an instructional team.

This research took place at San Jos\'{e} State University (SJSU), a Minority Serving Institution (MSI). The SJSU Learning Assistant Program serves over 5000 students annually in Physics, Chemistry, Biology, Math, and Computer Science courses. At this institution, LA supported classes include project-based courses, labs, and interactive lectures. 

\section{RIOT 2 Development and Data Collection}

Our team first conducted observations of LA-supported classes using the original RIOT. In some observations, we focused on observing the course instructor, and in others we focused on observing an LA. As we encountered situations where there was no button to describe what was happening, we made field note entries of what was happening. These new behaviors all were interactions between the person being observed and another member of the instructional team. For example, when we were observing an LA, there was no button for when that LA was watching the instructor lecture, asking the instructor a question, or dialoguing with the instructor. Likewise, when we were observing instructors, there was no button for when they were asking LAs questions, giving LAs instructions, or watching the LA address the whole class. 

RIOT has been hosted on other platforms previously, including FileMaker Pro, the UC Davis Generalized Observation and Reflection Platform (GORP), and through Google App Engine \cite{paul2018using}. Our team re-implemented RIOT on Google App Engine using Python and Google Cloud Platform for iterative development of RIOT 2. Our team iteratively added buttons and tested the implementation of our new protocol. 
\begin{figure*}
  \includegraphics[width=.83\linewidth]{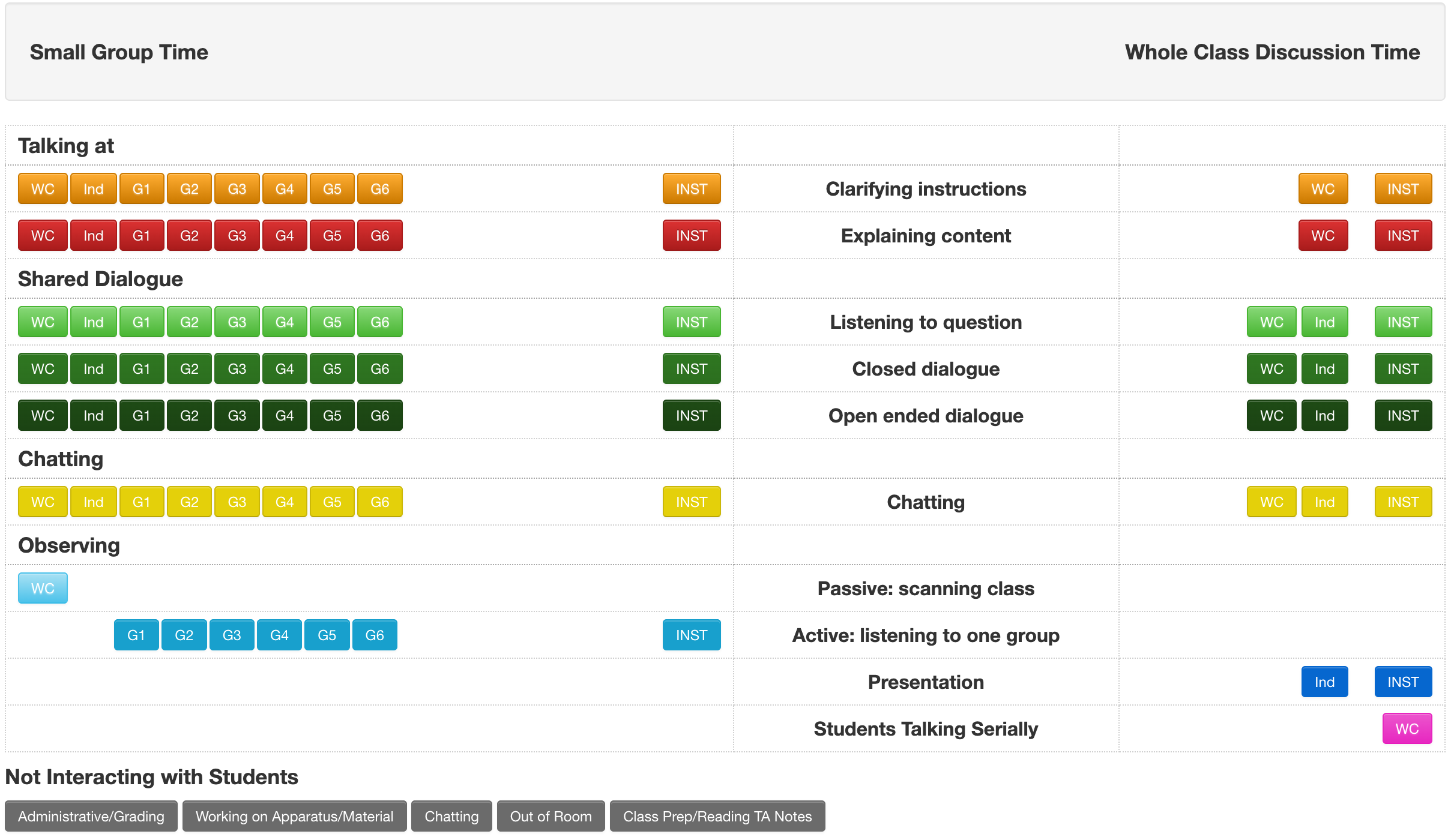}
  \caption{The interface of the RIOT 2, which an observer can use to observe an instructor or LA. \label{fig:riot2interface}}
\end{figure*}

Figure 2 depicts our updated RIOT 2. The major change was to add ``INST'' columns to both the left and right hand sides of the interface. The rows (categories of interaction) are unchanged. These protocol updates also required us to update how we presented plots. Because the time spent in instructor-instructor interactions was small in every class, we grouped all instructor-instructor interactions into one color (purple). We also moved ``Chatting'' to its own row because we noticed LAs chatting with students to build rapport.

To give an example of the use of the new buttons, consider a fictional scenario: while the LA is listening to the instructor lecture about a topic, the instructor asks the LA whether they had an alternative explanation to share with the class, and the LA shares their idea. Someone observing the LA would click \textbf{``INST'' Presentation} while the LA is watching the instructor, then \textbf{``INST'' Listening to question} while the LA listens to the instructor's question, then \textbf{``WC'' Explaining content} while they LA explains content to the class. 

Over the course of RIOT 2 development, our team of four observers has conducted 30 observations, where a single ``observation'' is any unique combination of date, observee, and observer. We note that two different observers watching the same instructor during the same class period is counted as two observations. Many of our observations are unusable due to coding bugs (e.g., an error in saving data) or because they used an older version of the tool. Eleven of these observations have been conducted using the updated RIOT 2 shown in Figure 2. Of these eleven observations, we have three instances of ``calibration'' where two observers observed the same instructor during the same class period for the purposes of establishing consistency across observers. We also have three instances of LA-faculty comparison, where an instructor and an LA were observed by different observers simultaneously. 

\section{Illustrative Examples}
We now present example RIOT 2 data. While these data are not intended to be representative, we choose these to show what our initial data look like, and to be a launching point for discussion about the potential of RIOT 2. 

\begin{figure}
  \includegraphics[width=\linewidth]{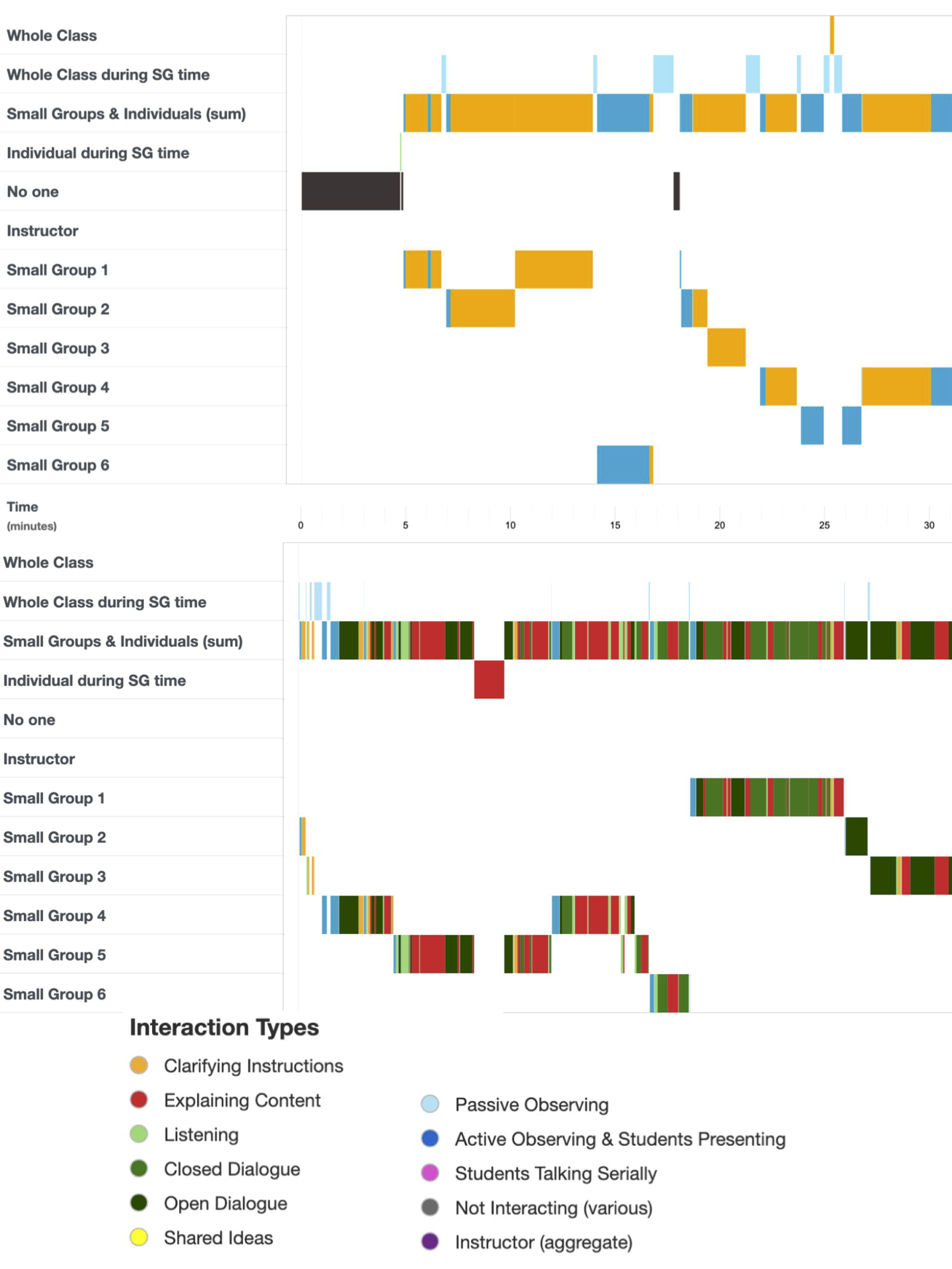}
  \caption{Simultaneous comparison of instructor (top) and LA (bottom) during one class period. The behavior key is at the bottom. \label{fig:comparison}}
\end{figure}

Some of our data are pairs of simultaneous observations of an LA and an instructor. In these cases, there are two observers in the room, with one observing the LA and one observing the instructor. Figure \ref{fig:comparison} depicts a 30 minute period of observation of a physics class working in small groups. The top depicts the instructor and the bottom depicts the LA. The timeline reads from left to right, and shows the amount of time each button (behavior) was active. The row labeled ``Small Groups \& Individuals (sum)'' is the sum of all small group and individual interactions during small group time, whereas the last 6 rows show this information broken down by group. This class is an interactive class in which students spend long chunks of time in small groups discussing activities at whiteboards. For the purposes of this paper, we omitted data from the earlier part of the class where the two observers were gathering calibration data.

The RIOT 2 reveals differences in LA and instructor behaviors in the same time interval. The instructor's time is primarily orange (clarifying instructions) and blue (observing students). On the other hand, the LA's time is primarily red (explaining content) and green (open and closed dialogue). The instructor spends more time observing students compared to the LA. The RIOT 2 output also gives a sense for the trajectory of the LA and instructor through the classroom over time. In the 30-minute segment, the LA and instructor rotate through the entire class with different starting points. All six groups interact with both the instructor and LA.

One notable feature of this example is that ``dialogue'' is primarily an LA activity; this is consistent with our other data, which suggest that LAs tend to engage in more dialogue than their faculty partners. This may be connected to the pedagogical training which explicitly emphasizes promoting dialogue through open and closed discussions \cite{van1997using}. We find it interesting that the instructor is primarily discussing instructions during this period; perhaps the instructor feels more responsible for ensuring the task at hand is clear, whereas the LA is more able to engage with student ideas. 

 \begin{figure}
  \includegraphics[width=\linewidth]{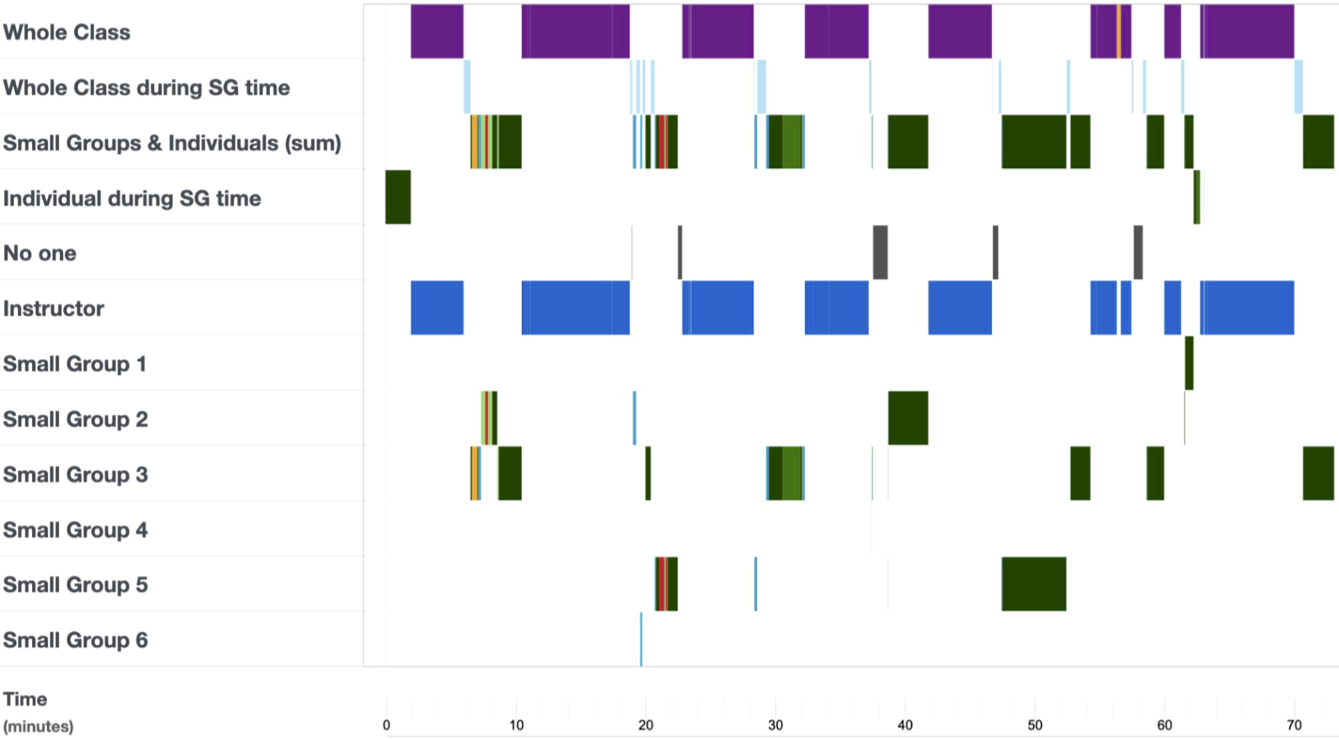}
  \caption{Observation of an LA during an interactive lecture. The color key is the same as Figure \ref{fig:comparison}.\label{fig:intlec}}
\end{figure}

As a contrast, Figure \ref{fig:intlec} depicts a 60-minute segment of a different, interactive lecture physics class. The data show the LA spending most of their time listening to the instructor lecture, with short discussions in small-groups interspersed. The LA spends almost their entire talk time in dialogue with the students.

Comparing the LA in Figure \ref{fig:intlec} with the LA in Figure \ref{fig:comparison}, we notice that even though both LAs emphasize dialogue, the LA in Figure \ref{fig:comparison} does significantly more explaining in this time period. We can also see how differences in the structures of the two class sessions come out in the RIOT 2 data. In Figure \ref{fig:intlec}, the LA only has 5-10 minutes at a time where they are interacting with students, and these are occurring between lecture segments. In Figure \ref{fig:comparison}, the students are working on an extended activity, in which they are in small groups for 30 minutes. 

\section{Discussion}
In this paper, we discussed how our research team developed the RIOT 2, an adaptation of the RIOT to be compatible with LA-supported classrooms. We describe changes to the original RIOT to account for interactions between members of the instructional team. We then shared several data excerpts to illustrate what the RIOT 2 data look like, and present some preliminary interpretations of that data. Even though the RIOT 2 was developed to account for classrooms with LAs, we argue that it can be utilized in any classroom with multiple members of an instructional team. 

As with any data collection approach, there are advantages and drawbacks to using the RIOT and RIOT 2 to collect observational data of classroom settings. The RIOT is time-efficient to use; training an observer can happen in an hour or less. The data are also classified and categorized in real-time, which is faster than collecting and analyzing video data. The drawback of this approach is that the RIOT does not allow for repeated categorization of the same class, which would be beneficial for validation. Our team has done some calibration of reviewers (to discuss discrepancies between interpretations of the categories), which we will discuss in future work. Finally, as the RIOT only stores information about what the instructor is doing, only instructor and LA consent is needed, since no identifying student data are recorded. This can be beneficial for observations of large classes, where it may be difficult to consent individual students for video analysis.

We emphasize that the data shown here are not meant to be generalizable of a given curriculum, LA, or instructor. They show how an instructor's time is used over a single class session (or segment of class), which may be similar to---or different from---other sessions of the same class. These individual classroom instances can help us identify interesting features that we could look for in a broader data set.

We plan to continue conducting extensive data collection to see what kinds of patterns can emerge across instructors, and across multiple class sessions of the same LA-faculty pairings over time. We are especially interested in how LA-faculty partnerships \cite{indukuri2022characterizing} might mediate LA and faculty behaviors in the classroom. Our team also plans to improve the reliability of our data collection by conducting additional calibration of observers and triangulating our data with videorecordings. 

The RIOT 2 may be of use for researchers of LA programs as well as instructors. Researchers can use the RIOT 2 to measure the presence (or absence) of active learning teacher moves in classrooms, track changes in teaching over time, or compare faculty. Similar to West and Paul \cite{west2013variation}, the RIOT 2 can support instructors and LAs in checking for consistency between their behaviors with their instructional goals related to active learning. It may also be helpful to compare different instructors teaching the same curriculum or coordinated class. 

RIOT 2 is available for use at \href{https://sjsuriot2-dot-sjsuriot.appspot.com/}{https://sjsuriot2-dot-sjsuriot.appspot.com/}. Our team also plans to make the source code available for others to host their own version, as well as edit their own buttons. For questions related to RIOT 2 use, please contact \href{mailto:gina.quan@sjsu.edu}{gina.quan@sjsu.edu}.

\acknowledgments{The authors thank Yashraj Bains and Leah Lira for their previous work on the RIOT 2. This material is based upon work supported by the National Science Foundation under Grant No. 2234071.}

 \bibliography{PERC2018} 

\end{document}